Title: Formational bounds of link prediction in collaboration networks

Authors: Jinseok Kim and Jana Diesner

Author 1) Jinseok Kim

> Institute for Research on Innovation and Science, Survey Research Center, Institute for Social Research, University of Michigan
>
> 330 Packard Street, Ann Arbor, MI U.S.A. 48105
>
> jinseokk@umich.edu

Author 2) Jana Diesner

> School of Information Sciences, University of Illinois at Urbana-Champaign
>
> 501 E. Daniel St., Champaign, IL U.S.A. 61820
>
> jdiesner@illinois.edu

Corresponding Author:

Jinseok Kim; jinseokk@umich.edu, 1-(734)-763-4994

Abstract

Link prediction in collaboration networks is often solved by identifying structural properties of existing nodes that are disconnected at one point in time, and that share a link later on. The maximally possible recall rate or upper bound of this approach's success is capped by the proportion of links that are formed among existing nodes embedded in these properties. Consequentially, sustained ties as well as links that involve one or two new network participants are typically not predicted. The purpose of this study is to highlight formational constraints that need to be considered to increase the practical value of link prediction methods for collaboration networks. In this study, we identify the distribution of basic link formation types based on four large-scale, over-time collaboration networks, showing that current link predictors can maximally anticipate around 25% of links that involve at least one prior network member. This implies that for collaboration networks, increasing the accuracy of computational link prediction solutions may not be a reasonable goal when the ratio of collaboration ties that are eligible to the classic link prediction process is low.

# Introduction & Background

To what extent can we predict which pair of nodes will form a link in a social network? Being able to answer this question helps with a wide range of practical applications such as recommender systems (Resnick & Varian, 1997) and collaborative filtering (Yan & Guns, 2014), or the planning of interventions to prevent the spreading of diseases and rumors (D.-B. Chen, Xiao, & Zeng, 2014). Network scientists refer to this task as link prediction.

Computational approaches to this problem usually focus on finding strong predictors for links in a future network based on structural properties of the same network at an earlier point in time (Lü & Zhou, 2011; Taskar, Wong, Abbeel, & Koller, 2003). Many solutions to this task have been presented, especially for collaboration networks (e.g., H. Chen, Li, & Huang, 2005; Guns & Rousseau, 2014; Liben-Nowell & Kleinberg, 2007; Yan & Guns, 2014). Features considered in these studies are based on theories, empirical observations, and assumptions. For example, a main feature considered by the widely used Common Neighbors predictor is the number of common alters of two nodes, which is supported by the observation that people who have friends or coauthors in common are more likely to connect with each other in the future (Kim & Diesner, 2017; Liben-Nowell & Kleinberg, 2007; Mollenhorst, Volker, & Flap, 2011).

Most link predictors are trained on data where two nodes are present but not linked at time $t_n$ (past network), and share a link at time $t_{n+1}$ (present network). The recall rate (coverage) or upper bound of the potential successfulness of this approach is capped or defined by the proportion of links in networks that are formed according to this principle. This common practice does typically account for (a) sustained links, and also not for (b) links in which at least one of the two nodes is a new network participant at $t_{n+1}$. Correctly anticipating case (b) would be extremely hard for the outlined link prediction techniques as new nodes do not yet have a structural history and hence no prior information about them that can be leveraged for prediction. This division of link formation processes into basic types leads to the following question. How many links in a network are formed between previously disconnected members, how many links sustain over time, and how many links involve one or even two new nodes? This chain of thought leads to another question about the applicability of link predictors: To what extent can high-performing predictors provide comprehensive (recall) and reliable (precision) knowledge about link formation in a network? Answers to these questions can help us to identify limitations of current link predictors and suggest how to overcome these limitations. To answer these questions, the number of links of each aforesaid type first needs to be counted. For the domain of coauthor networks, scholars have categorized coauthors into several types. For example, according to a series of coauthorship studies (Braun, Glanzel, & Schubert, 2001; Price & Gürsey, 1976; Schubert & Glanzel, 1991), such networks are composed of a group of core authors who continue to publish, and their coauthors who are categorized as one of "newcomers," "terminators," and "transients." Among these coauthor types, a recent study of the life-long careers of 3,860 computer scientists found that 72.7% of coauthors are transient, i.e., collaborating once but never again with the same scientists (Cabanac, Hubert, & Milard, 2015). However, such observations have rarely been connected to the constraints and applicability of link prediction in coauthor networks.

In this paper, we present answers the questions raised above. This work complements prior link prediction studies by measuring the distribution of the types of link formation in collaboration networks, and tests how many of those links can be explained or predicted with high-performing link predictors. For this purpose, we analyze four large-scale, over-time collaboration networks from the fields of biomedicine, computer and information science, physics, and a nation. We next introduce our datasets and describe how we prepared them for analysis. We then describe our methodology, present our findings, and discuss implications and limitations of our work.

## Data

In coauthor networks, two authors (nodes) are connected by links if they publish papers together. Here, we ignore the direction of links (undirected graphs) and only care about the existence of links between nodes (unweighted or binary graphs). Coauthor networks have been frequently used for link prediction studies. This might be partially because they can be easily constructed, even at a large scale, from publicly available bibliographic data. Another reason might be that bibliographic records include temporal information on publications, which is essential for some link prediction techniques. In the following section, the datasets used in this study are described in detail.

We analyze four over-time, large-scale publication datasets, which were derived from MEDLINE (domain of biomedicine), DBLP (computer science and informatics), APS (Physics), and KISTI (country-level data for Korea). Table I summarizes key features of these datasets.

One critical feature of scholarly publication data is the accuracy of entity resolution in terms of splitting up author names that are shared by multiple people into as many nodes and merging variations in referring to individuals (Fegley & Torvik, 2013; Kim & Diesner, 2016). Our original DBLP and KISTI are already disambiguated (for details on KISTI, see Kim, Tao, Lee, & Diesner, 2016; for details on DBLP, see Reitz & Hoffmann, 2011). For MEDLINE and APS, we obtained algorithmically disambiguated data as outlined below.

*Table 1: Overview of Datasets*

| Scope | Dataset | Time Frame | No. of Papers |
|---|---|---|---|
| Biomedicine | MEDLINE | 1991-2009 | 342,158 |
| Computer Science | DBLP | 1991-2009 | 231,161 |
| Physics | APS | 1991-2009 | 241,329 |
| Articles published in Korea | KISTI | 1991-2009 | 273,869 |

*MEDLINE*

MEDLINE is the bibliographic database of the National Library of Medicine[1]. This digital library is composed of journal papers from the fields of biology and medicine from the year of 1950 forward. For each paper, the following information is indexed (if available): unique identifier (PMID), name and affiliation per author, paper title, journal, and a selection of predefined keywords, which are referred to as medical subject headings (MeSH). The Author-ity 2009 dataset (Torvik & Smalheiser, 2009) contains algorithmically disambiguated author names in MEDLINE up to the year of 2009 (Author-ity 2009). In Author-ity 2009, author name ambiguity is resolved with up to 98~99% accuracy through a similarity calculation based on the names of authors and coauthors, paper title, journal name, and MeSH terms (Lerchenmueller & Sorenson, 2016; Torvik & Smalheiser, 2009). To create a subset of the MEDLINE data that is similar in size to the other datasets considered in this study, we extracted the articles with the MeSH term 'brain', which is one of the most frequently occurring MeSH terms (Newman, Karimi, & Cavedon, 2009).

---
[1] https://www.nlm.nih.gov/bsd/licensee/medpmmenu.html

*DBLP*

DBLP (Digital Bibliography & Library Project)[2] contains records of more than four million papers from journals and conferences in computer science and information science starting from the 1950s onwards. In this database, author names have been disambiguated by computing similarity scores based on author names, coauthorship information, and manual post-processing of incorrect data points reported by end-users (Reitz & Hoffmann, 2011). Author name disambiguation in DBLP is reported to be highly accurate (around pairwise F1 = 0.90 and above), even beating many advanced algorithms when tested on multiple labeled datasets (Kim, 2018; Kim & Diesner, 2015). For this study, we selected a subset of papers published in 392 computer science journals[3].

*APS*

The American Physical Society (APS) provides publication information on the papers that appeared in the Physical Review series[4], which entails prime journals in physics, from 1893 onward. In the original data, author names are not disambiguated. We disambiguated the data by implementing the same algorithm described in Martin, Ball, Karrer, and Newman (2013)[5], which considers the name of authors and coauthors, affiliation, and venue. The following error rates are reported for this algorithm: incorrect merging of 3%, and erroneous splitting of 12% of a sampled set of authors (Martin et al., 2013).

*KISTI*

This dataset was collected, disambiguated, and provided by the Korea Institute of Science and Technology Information (KISTI)[6]. KISTI indexes conference proceedings and journal papers that have been published in Korea from the late 1940s to today. Author names in these data were distinguished by using a clustering algorithm considering features including surface form of names, affiliation, coauthor, paper title, and name of the journal or conference (Kim et al., 2016). The accuracy of this automated solution is reported to be 0.94 (pairwise F1). Further inspections and corrections were performed by KISTI's human experts. For this study, we only consider journal papers in order to be consistent with prior link prediction studies.

Finally, for each of the four datasets, we counted the number of authors per paper, and excluded papers with an exceedingly large number of coauthors as they could have an un-proportionally large impact on t results. The following thresholds were defined for each dataset, and papers with more authors than the thresholds were excluded from analysis: DBLP (7), APS (14), MEDLINE (14), and KISTI (8). The resulting datasets still contain 98%~99% of all papers from our original datasets.

Method

*Time-Sliced Network Construction*

Link prediction typically involves the following steps: First, ~~the~~ network data are divided into two networks: one portion for predicting links (e.g., a network for 2000-2004 period, hereafter past network), and another portion for evaluating the accuracy of the prediction (e.g., a network for the year 2005, hereafter present network). Then, the set of nodes that appear in both networks (hereafter, existing nodes) is identified (typically based on matching node identities). For the existing nodes, linkage in the present network is often predicted based on network properties in the past network. For example, a disconnected pair of nodes in the past network that is connected to the same *N* alters (where *N* is a threshold value) is assumed to form a link in the present network. The outcomes of this process are then evaluated by checking whether the predicted links actually appear in the present network.

---

[2] http://dblp.org/xml/release/; for this study, we downloaded the April 2015 release.
[3] A list of 392 journal was obtained from Thomson Reuters Journal Citation Report 2012 for the category "Computer Science". We retrieved records on these papers published in these journals from DBLP.
[4] http://journals.aps.org/datasets; for this study, we obtained the APS 2014 release version under the permission of the American Physical Society.
[5] Mark E. J. Newman at the University of Michigan Department of Physics kindly provided the disambiguation code
[6] http://scholar.ndsl.kr/index.do; for this study, we obtained the KISTI 2016 version under a research agreement with the Korea Institute for Science and Technology Information.

Following this common procedure, we also divided our network data into past and present networks, and retrieved the nodes that appear in both networks. For papers with more than two authors, every entailed author pair is considered as a separate link. For example, a paper written by three authors A, B, and C generates three collaboration links of A-B, A-C, and B-C. We sliced each of the four datasets – MEDLINE, DBLP, APS, and KISTI - into past and present networks. To see how different time frames affect results, we varied time frames of past and present networks, following the practice of previous studies on this topic (e.g., Choudhury & Uddin, 2017, 2018; Liben-Nowell & Kleinberg, 2007). For example, papers in each dataset were divided into two subsets containing papers published during past (e.g., 1991~1995; 5 years) and present (e.g., 1996; 1 year) periods, respectively, for network construction. This slicing was repeated over different combinations of past and present periods using 1, 3, and 5 years as units; resulting in a total of 9 past-present time frames for a dataset: [past 1 year | present 1 year], [past 1 year | present 3 years], [past 1 year | present 5 years], [past 3 years | present 1 year], [past 3 years | present 3 years], [past 3 years | present 5 years], [past 5 years | present 1 year], [past 5 years | present 3 year], and [past 5 years | present 5 years]. In addition, we repeated our measurements with sliding windows. For example, for the [past 5 years | present 3 years] time frame, we applied the same measurements for the past 1991~1995 period (5 years) and the present 1996~1998 period (3 years), the past 1992~1996 period (5 years) and the present 1997~1999 period (3 years), and so on. This was done to see if our measurement results and any particular patterns depend on specific windows.

*Link Types for Analysis*

Unlike prior studies, we do not estimate but measure (i.e., count) the rate of suggested links based on the past network that actually occur in the present network. More specifically, we identify all links in the present networks that involve at least one of the existing nodes. Theoretically, four types of links are possible in an evolving collaboration network, which are illustrated in Table II. In that table, X and Y represent nodes that appear both in the past and present networks, and Z and W denote "newcomers" in the present network. A dash between nodes represents a link, while a comma indicates the absence of a link.

*Table 2: Summary of Link Types for Analysis*

| Type | Past Network | Present Network | Considered for Analysis | Interpretation | Predictability |
|---|---|---|---|---|---|
| **A** | X−Y | X−Y | Yes | Sustained collaboration b/w existing authors | Possible Typically NOT target of prediction |
| **B** | X, Y | X−Y | Yes | New collaboration b/w existing authors | Typically target of prediction |
| **C** | X, Y or X−Y | X−Z or Y−W | Yes | New collaboration b/w existing and new authors | Difficult Typically NOT target of prediction |
| **D** | None | Z−W | No | New collaboration b/w new authors | Even more difficult Typically NOT target of prediction |

First, a link between nodes X and Y that occurs in both the past and the present network represents an instance of sustained or continued collaboration (hereafter Type A). Second, nodes X and Y being present but disconnected in the past network and forming a link in the present network represent a new collaboration between existing scholars (hereafter Type B). Third, a link from a node in the past network (X or Y) with a new node Z in the present network represents a new collaboration between partially old (X or Y) and partially new (Z) scholars (hereafter Type C). Last, two new network participants (appearing in the present network, but not in the past network) could form a link (hereafter Type D). As we focus on links that involve at least one existing nodes, and new links between new members (Type D) are extremely difficult to anticipate, we only consider links of types A, B, and C for the results section.

Results

Bounds for Predictable Link Types

The ratios of types A, B, and C in each of four datasets are reported in Figure 1~4. Each figure has 9 subfigures reporting results for each type measured over 9 different time frames with a yearly sliding window. For example, the upper-left subfigure in Figure 1, titled 'MEDLINE: Type A (Past 1 Year),' shows the ratio changes ($y$-axis) of Type A over Type A + B + C that were measured for the 'past 1 year' (1) with three different present year ranges: 1 year (circles), 3 years (triangles), and 5 years (x-crosses), and (2) with each starting present year moving at a yearly resolution ($x$-axis). Note that triangle-shaped data points for the 'present 3 years' stopped at 2007 (in other words, 2008 and 2009 results are not reported). This is because the 'present 3 years' for 2008- and 2009-year windows could not be extended to 2010 and 2011 as our datasets contain records of publications published from 1991 up to 2009. This also explains why the data points for the 'present 5 years' stopped at 2005: the 2005~2009 period is covered in our data, but its subsequent periods (2006~2010, 2007~2011, and so on) are incomplete due to missing information after 2009.

Results in Figure 1~4 show that the ratios of links of Type A (sustained collaboration) are around 10~30% of the sum of Type A + B + C (see left subfigures in Figure 1~4). Two observations are worth noting. First, this ratio do not fluctuate much over time windows within each past-present network period across all four datasets. This means that the tendency of authors to work with previous collaborators has not changed much. Another notable observation is that the x-crosses plots appear below the triangles plots, which in turn appear below the circles plots. In other words, if a past network period is shorter and its paired present network period is longer, the ratio of Type A links in the past-present network pair is lower than the longer past and shorter present network pairs. This observation implies that as we extend the present network periods in our analysis, we find more collaboration instances categorized as other than Type A, thus decreasing the reported ratios of Type A. What this means is that continued collaborations among existing authors tend to be captured disproportionally more often in shorter present networks. A possible explanation for this observation might be that prior coauthoring experience improves communication and resource allocations among continued coauthors, leading to subsequent research papers being produced faster than those resulting from new collaborations between existing authors or existing authors and newcomers[7].

---

[7] This demonstrates why varying past-present network time frames matters for this study. The idea of using different past-present network periods was suggested by one of the reviewers of this paper.

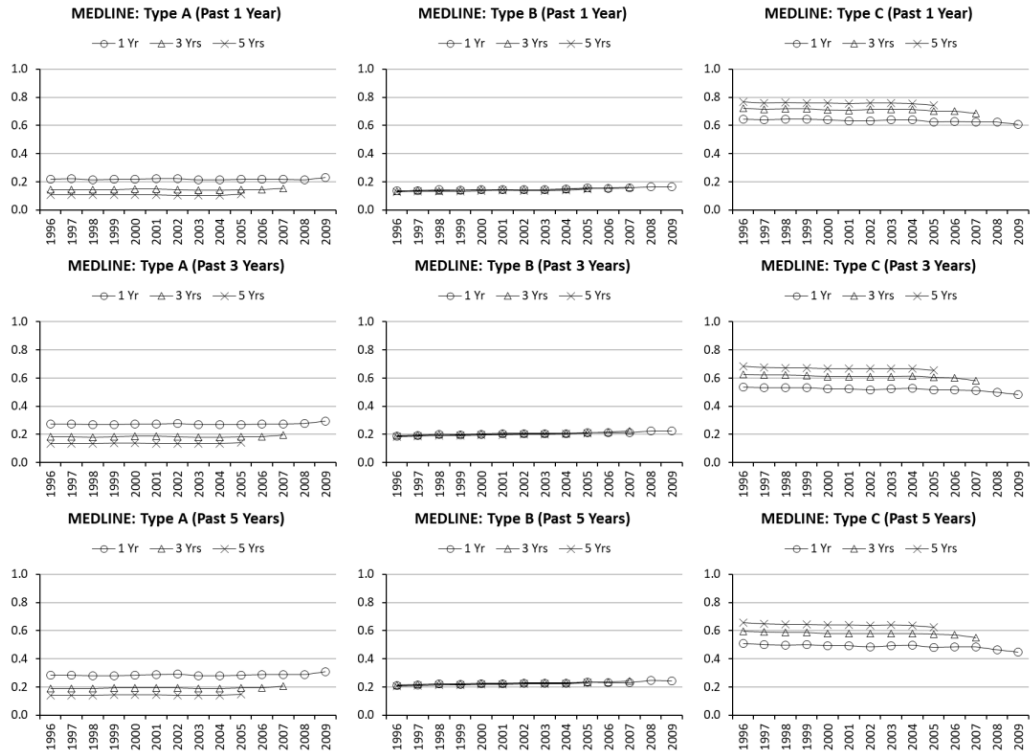

*Figure 1: Over-Time Ratio of Link Types in MEDLINE (Type A: sustained collaborations between existing nodes, Type B: new links between existing nodes, Type C: links between existing nodes and new nodes)*

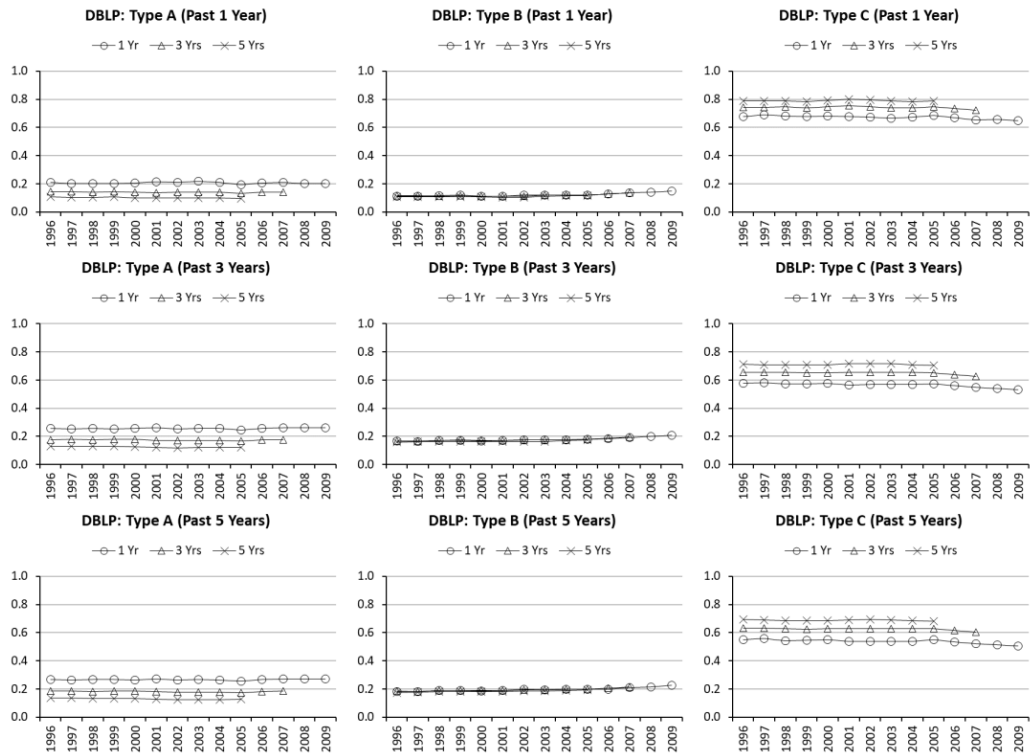

*Figure 2: Over-Time Ratio of Link Types in DBLP (Type A: sustained collaborations between existing nodes, Type B: new links between existing nodes, Type C: links between existing nodes and new nodes)*

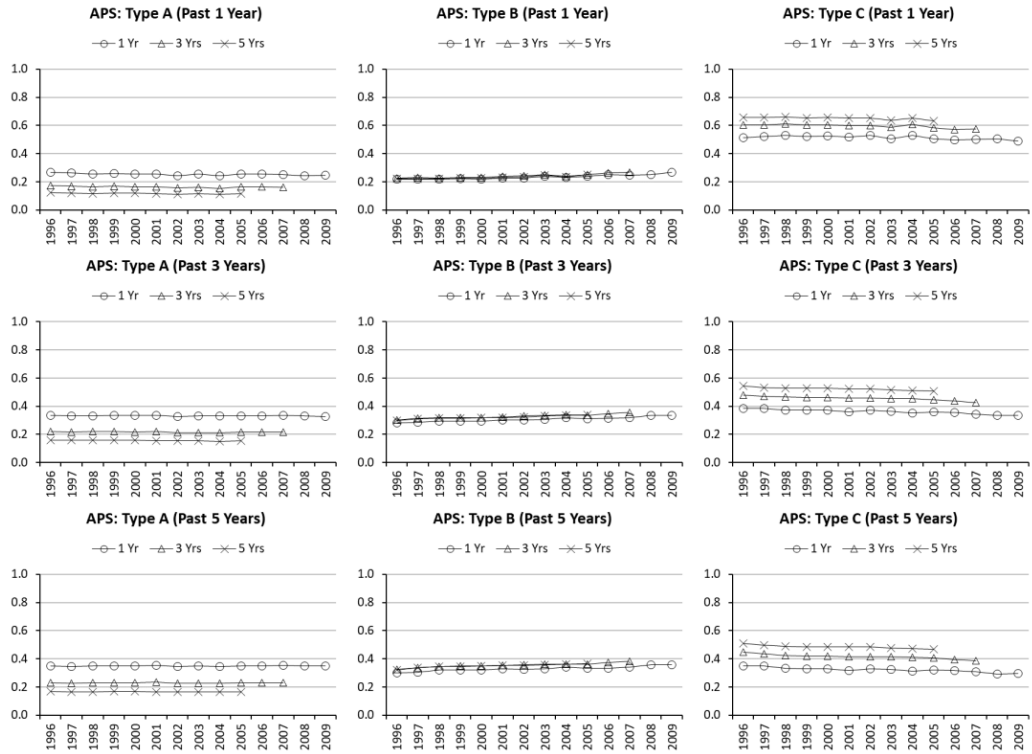

*Figure 3: Over-Time Ratio of Link Types in APS (Type A: sustained collaborations between existing nodes, Type B: new links between existing nodes, Type C: links between existing nodes and new nodes)*

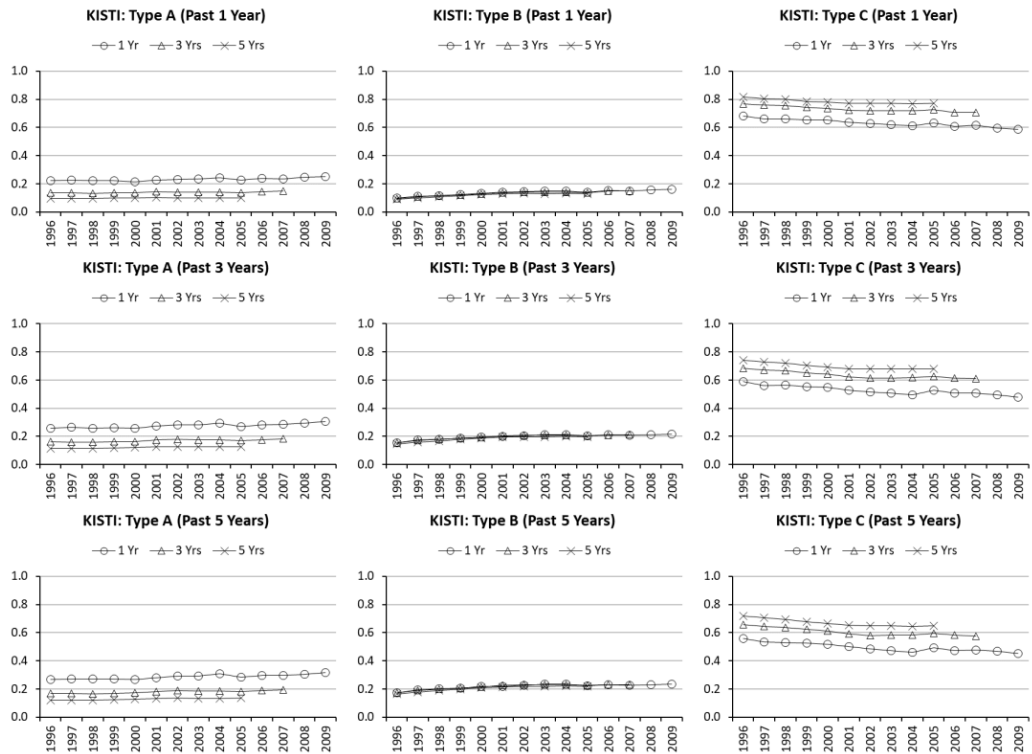

*Figure 4: Over-Time Ratio of Link Types in KISTI (Type A: sustained collaborations between existing nodes, Type B: new links between existing nodes, Type C: links between existing nodes and new nodes)*

With regard to Type B links, which are the focus of many link prediction studies, the results in Figure 1~4 reveal that these links constitute 10~25% (MEDLINE, DBLP, and KISTI) or 20~30% (APS) of the sum of Type A + B + C links, and that their ratios slightly increase regardless of the past-present network periods in each dataset. This is shown by the x-crosses, triangles, and circles plots that overlap and move towards the upper-right in very low slopes (see middle subfigures in Figure 1~4). This implies the caveat that, even if links can be predicted with high recall and precision, the prediction will account for less than 25% (MEDLINE, DBLP, and KISTI) or 30% (APS) of coauthorship relations involving at least on existing node (i.e., against Type A + B + C).

When compared with the ratios of Type A links, Type B links are slightly more or less frequent than Type A links across most time-windows in MEDLINE, DBLP, and KISTI. In APS, Type B is quite similar to or slightly more frequent than Type A. This means that roughly speaking, more or less than 50% of the links among existing nodes (i.e., Type A + B) represented continued connections. In other words, authors are slightly more or less likely to write papers with their previous coauthors than with unfamiliar existing authors. While the Type A ratios vary depending on the past-present time slicing choices (left subfigures), the Type B ratios are not much affected by those choices (middle subfigures).

Finally, the plots in Figure 1~4 suggest that new collaborations between an existing and a new author (Type C) are the most common link type. Overall, Type C links account for 30% ~ 80% of all links per past-present network period and, like Type A, those ratios do not change much across all considered time windows. This observation indicates that authors commonly publish papers with new coauthoring partners. For example, new graduate students might write papers with their advisor, other graduate students, and later on with their post-doc advisor, then with their advisees or colleagues, and so on. Another notable observation is that contrary to Type A, the x-crosses plots appear above triangles and circles plots in the right subfigures in Figure 1~4. These trends imply that if a past network period is shorter and its paired present network period is longer, its ratio of Type C links is higher than the longer past and shorter present network pairs in each subfigure for Type C across all four datasets. What this tells us is that by extending the present network periods, new collaborations between existing and new authors are better detected than relying on shorter ones, which aligns well with the abovementioned interpretation that continued collaborations are more prominent in shorter present network periods. Despite such variations of Type C ratios are dependent on the past-present network periods, however, Type C links are found to be dominant in all datasets. Interestingly, the upper bounds of the Type C ratios, i.e., 68~80% in DBLP (see right subfigures in Figure 1), are comparable to the ratio of coauthors (72.7%) who have worked together only once for a sample of computer scientists across their publication careers as reported in Cabanac et al. (2015).

*Bounds for Link Predictor Performance*

Focusing only on Type B, we tested a commonly used, high-performing predictor, called Adamic-Adar, against datasets as sliced above into varied past-present network periods and sliding windows. This predictor is based on a similarity measure from Adamic and Adar (2003), which identifies two webpages to be more similar if they share features that are rarely shared by others than if they share features frequently observed for others as well.

In the following equation for the predictor, S($x$, $y$) is the prediction score for a pair of node $x$ and $y$, and $\Gamma(x)$ is the set of nodes connected to $x$. According to the equation (1), two nodes in a past network are more likely to form a link in a present network if they share neighboring nodes that are rarely connected to other nodes (i.e., low-degree nodes) than if they share neighboring nodes that are frequently connected to other nodes (i.e., high-degree nodes).

$$S(x,y) = \sum_{z \in \Gamma(x) \cap \Gamma(y)} \frac{1}{log|\Gamma(z)|} \qquad (1)$$

This concept was adopted by Liben-Nowell and Kleinberg (2007) and in following studies as a major link predictor. The Adamic-Adar predictor showed the highest accuracy among 9 predictors in Liben-Nowell and Kleinberg (2007). It has been also used as a baseline predictor in several studies for evaluating the performances of various predictors (e.g., Choudhury & Uddin, 2017, 2018; Guns & Rousseau, 2014). We used a Python package, *Linkpred* (Guns, 2014), to implement the Adamic-Adar predictor.

In Figures 5~8, the prediction results by the Adamic-Adar predictor are shown in recall-precision charts. Recall (*x*-axis) measures the retrieval rate of relevant items, and precision (*y*-axis) the accuracy within the set of retrieved items[8]. In the figures we show recall-precision plots for three selected frames in each past and present network period. For example, for the past 1 year & present 1 year frame (left-upper subfigure in Figure 5), results are reported for 1996 (blue solid line: past year = 1995, present year = 1996), 2002 (green dashed line: past year = 2001, present year = 2002), and 2009 (red dotted line: past year = 2008, present year = 2009). Those tree frames were selected by choosing the lower and upper bound years of present networks and the (approximately) middle year between them. The hyphened number (in grey color) in the center of each subfigure represents a pair of past (number before a hyphen) and present (number after a hyphen) network periods.

---

[8] In some fields, such as natural language processing, recall and precision are often inversely related and therefore an average score such as the F metric (e.g., harmonic mean of precision and recall) is calculated.

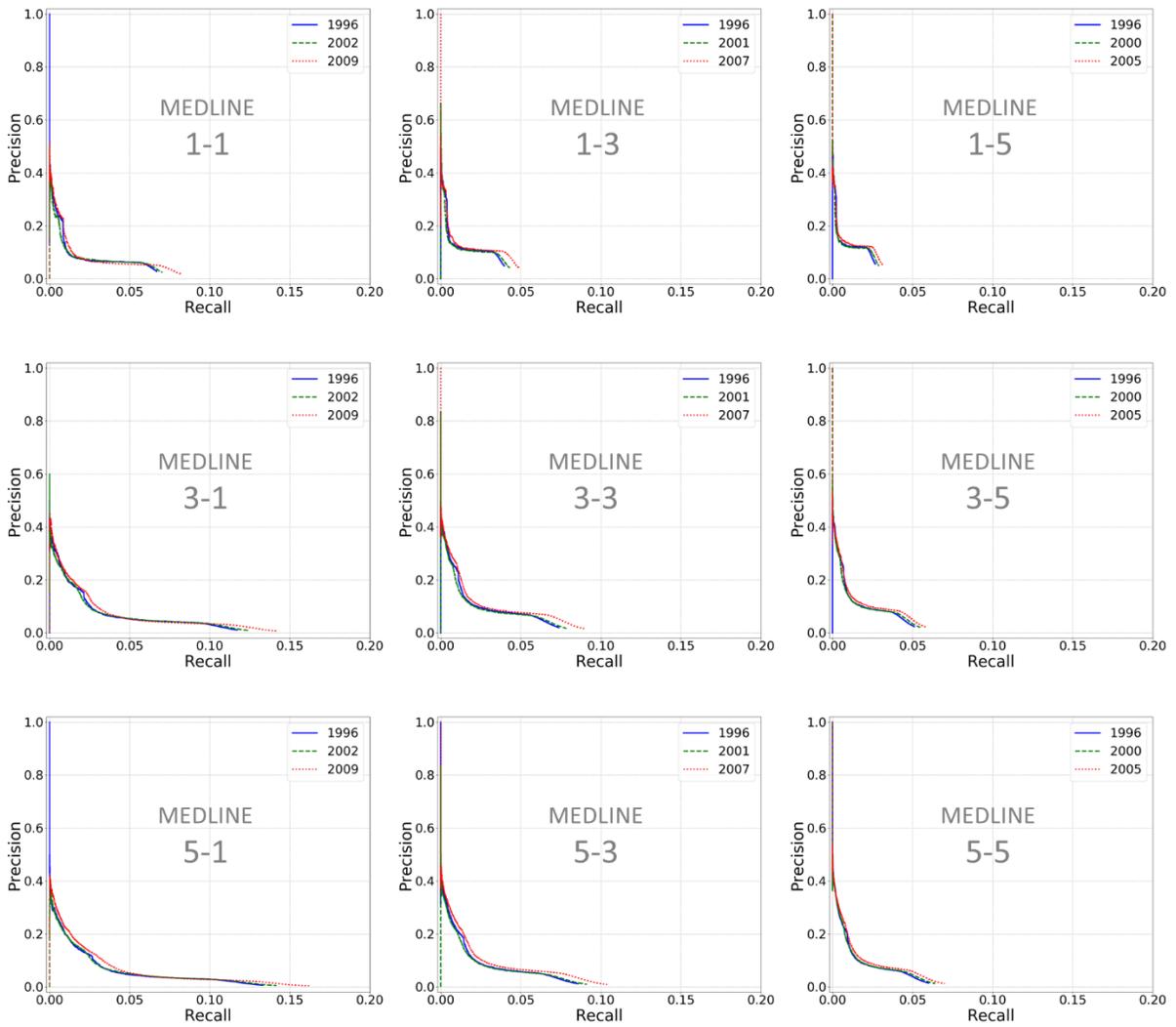

*Figure 5: Recall-Precision Curves for Link Prediction Performance by the Adamic-Adar Predictor on MEDLINE*

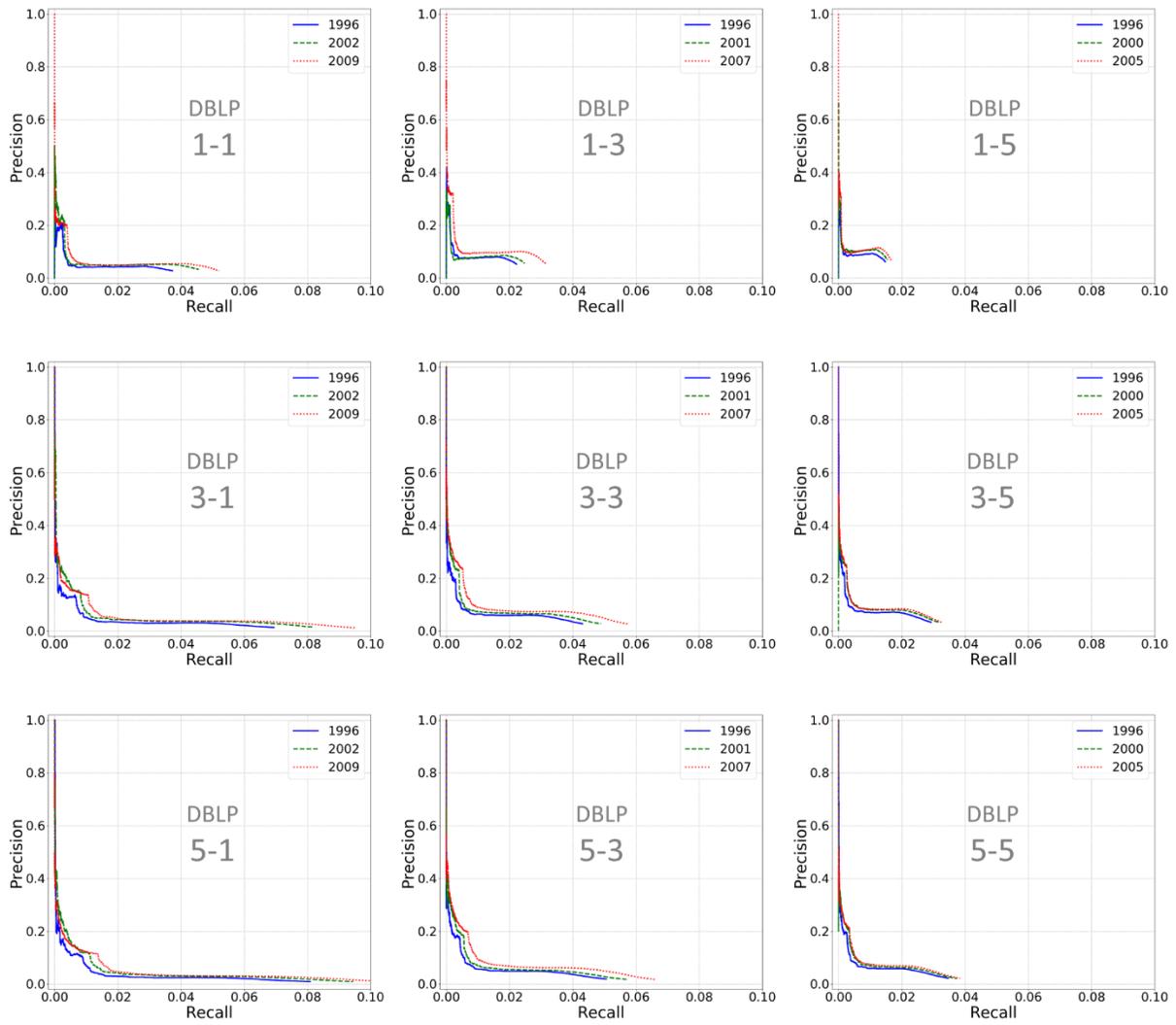

*Figure 6: Recall-Precision Curves for Link Prediction Performance by the Adamic-Adar Predictor on DBLP*

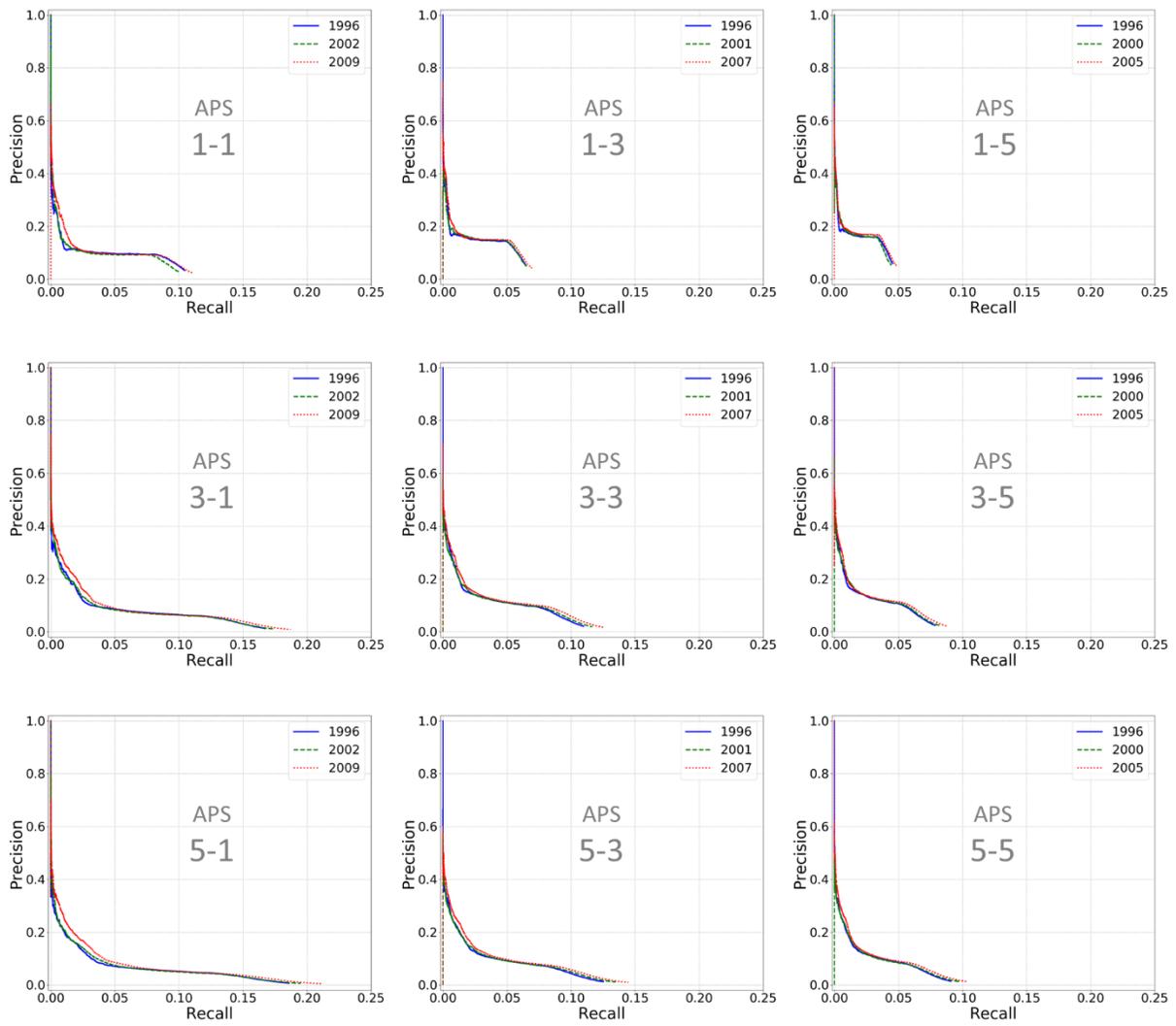

*Figure 7: Recall-Precision Curves for Link Prediction Performance by the Adamic-Adar Predictor on APS*

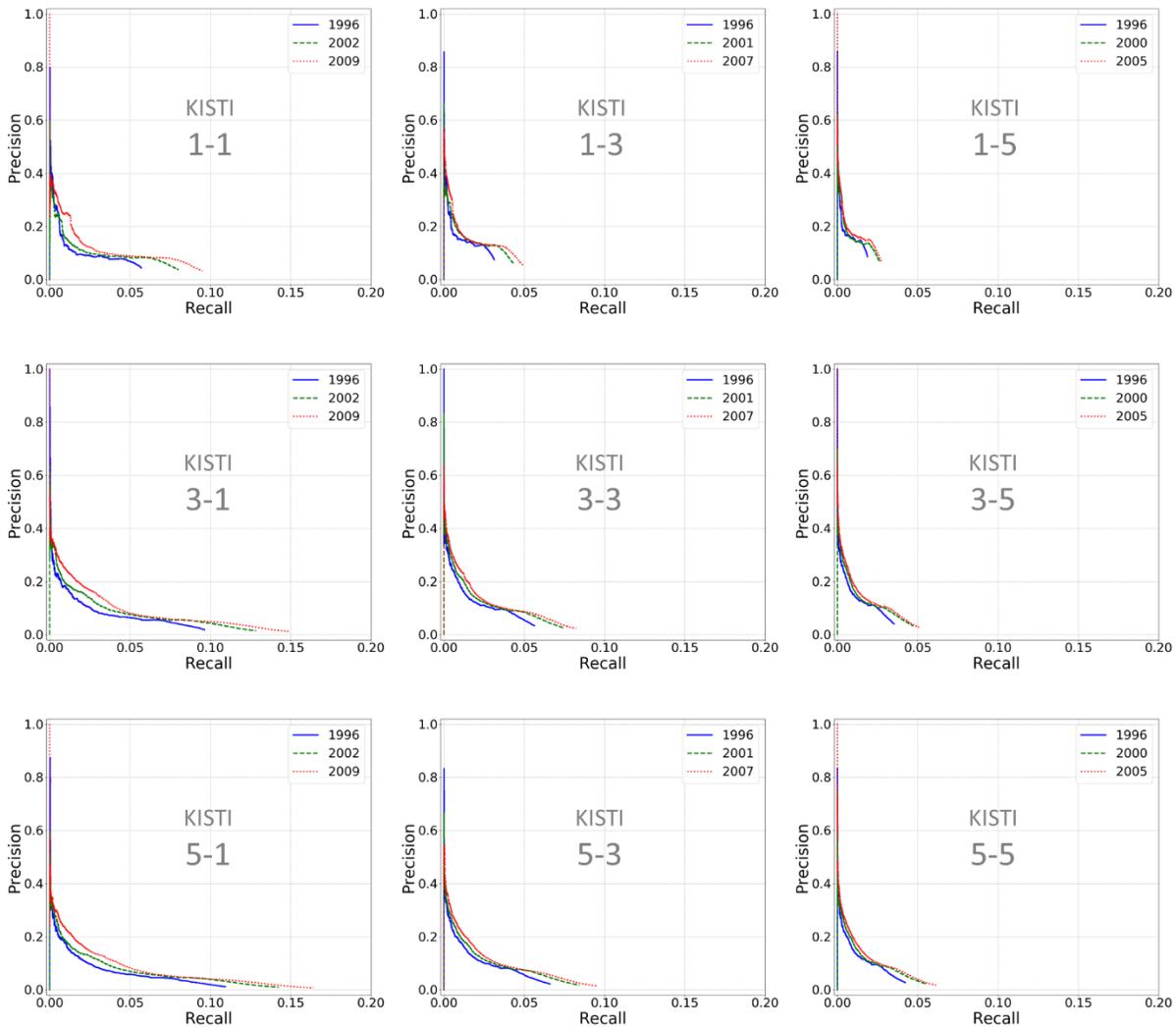

*Figure 8: Recall-Precision Curves for Link Prediction Performance by the Adamic-Adar Predictor on KISTI*

Figure 5~8 show Adamic-Adar-based recall rates of around 0.04~0.15 for MEDLINE, 0.015~0.1 for DBLP, 0.05~0.20 for APS, and 0.025~0.15 for KISTI. Precision rates increased up to 1.0 in all datasets but with extremely low recall rates. The tails of recall-precision plots that stretch towards the lower-right directions became shorter for the time-frames with longer present years. This means that as present years were extended longer, recall rates decreased across all past years. Although precision rates became slightly higher with longer present years, such increases were offset by the decreases of recall rates. This implies that as present network periods are extended, more Type B links between existing authors become identifiable, but the Adamic-Adar predictor fails to predict many of them. The low recall rates in Figures 5~8 mean that even one of the best performing link predictors for collaboration networks can detect only a small portion of new links between existing nodes in our datasets. These observations imply that classic link prediction methods can lead to incomplete knowledge about link formation mechanisms in large-scale collaboration networks. This is mainly due to the facts that (1) the common practice of link prediction studies only considers a particular type of links, i.e. links between existing nodes that were not connected in the past but are linked in the present network (Type B), (2) links of Type B account for only about a quarter of all links (Type A + B + C), and (3) imperfect prediction accuracy of technical solutions relying only on network structures further reduces link predictability.

An exception to this methodological practice in link prediction studies for collaboration networks is Mohdeb, Boubetra, and Charikhi (2016), who attempted to predict Type A links (sustained collaborations) using link strengths between two authors and their attributes, such as professorial ranks. Their study reported about 40% of accuracy in predicting whether existing collaboration links continue to persist, but did not consider links of types other than A in collaboration networks.

Bounds for Topological Prediction

Another notable exception is an extension that leverages the preferential attachment model and its variations (Barabási et al., 2002; Milojević, 2010; Newman, 2001a). According to this model, the probability for a node to form a link to node *x* is proportional to the number of nodes that are already linked to *x*. Several studies have used this model to predict Type B links (e.g., Degree Product in Liben-Nowell & Kleinberg, 2007)[9]. Unlike other predictors, however, the preferential attachment model is also used to predict link formation according to Types A and C.

When applied to all link types (A, B, and C), the prediction model's performance is not evaluated by recall and precision at a dyadic link level. This is because the preferential attachment model is stochastic (Perc, 2014). In other words, sets of linked nodes can be different even when the model is run repeatedly on the same data. Instead, previous preferential attachment studies tend to first detect the node degree distribution in a empirical target network, model the mechanism generating such a distribution, and then evaluate their proposed model by checking whether the node degree distribution of the model-generated network follows a pattern similar to that of the target network (e.g., Barabási et al., 2002; Pennock, Flake, Lawrence, Glover, & Giles, 2002). Especially, the model often aims to generate a network characterized by a specific network topology in which its node degree distribution follows a power-law pattern of $x^{-\alpha}$ (*x* is a node degree and $\alpha$ is a scaling parameter). This implies that the model's prediction holds true for the *x* range that exhibits the power-law distribution pattern.

To see whether link formation in our coauthor networks fits the preferential attachment model, we tested how many nodes in a network can be attributed to producing a power-law degree distribution. Technically, this means that we counted the number of nodes that have a nodal degree *x* that falls within the *x* range associated with a power-law degree distribution. If many nodes in a network are linked to each other in a way that generates a power-law obeying distribution of nodal degree, we may conjecture that links in the network are likely to be formed according to a power-law generation mechanism like preferential attachment[10]. For this task, we used an R package, *poweRlaw*[11], which implements the rigorous power-law fitting method proposed in Clauset, Shalizi, and Newman (2009). Figure 9 illustrates the process of deciding the ratios of nodes that generate the degree distributions following a power-law.

---

[9] For a detailed explanation for the Degree Product predictor, see Appendix.
[10] This does not mean that all preferential attachment models are designed to explain power-law obeying networks. However, many studies on preferential attachment have attempted to model power-law obeying networks.
[11] https://cran.r-project.org/web/packages/poweRlaw/index.html

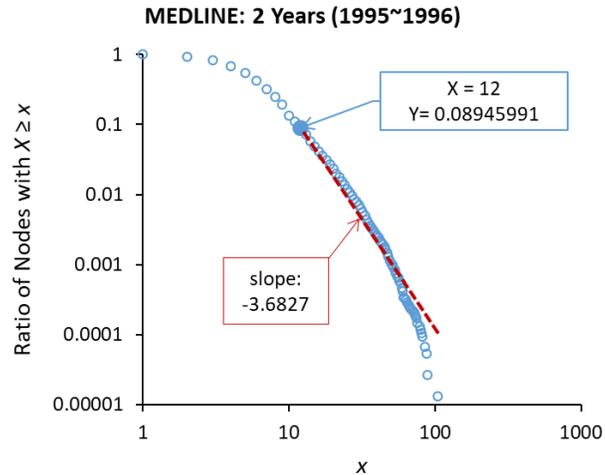

*Figure 9: An Illustration of Fitting a Power-Law Slope and Calculating the Ratio of Nodes with Minimum X or more*

In Figure 3, a degree distribution (blue circles) of an example network (MEDLINE: 2 Years (1995~1996)) is shown on a double logarithmic plot, where the value *x* on the horizontal axis is the node degree (i.e., number of coauthors per author), and its corresponding value on the vertical axis is the ratio (proportion) of the number of nodes (i.e., authors) that has *x* or more degrees over the total number of nodes. Using the fitting methods in Clauset et al. (2009), the distribution is fitted to a power-law distribution (red dotted line) with a slope of -3.6827 (parameter = 3.6827), but on a limited *x* range ($x \geq 12$). Here, we can obtain the ratio of nodes with *x* or more (Y = 0.08945991) using the minimum *x* (=12).

We repeated this fitting and calculation procedure on our four datasets. Unlike a typical link prediction method where data are divided into training (past network) and test (present network) sets, the power-law fitting was conducted on each longitudinal dataset (one per dataset). To study the effects of time frames, we sliced each dataset into 2, 6, and 10 years with a yearly sliding window for the 1996~2009 period. Figure 10 reports scaling parameters (left subfigures), minimum *x* values (*x-min*, middle subfigures), and ratios of nodes with *x-min* or more (right subfigures). A year on an *x*-axis represent the upper year of 2, 6, and 10-year time frames. For example, for the 2-year time frame, 2009 means the 2008~2009 period. Note that reporting for 10-year time frames starts from the year of 2000 because that year is the minimum starting year covering prior 10 years (1991~2000) in our datasets (see Table 1). Also note that in the left subfigures, filled shapes represent that the power-law fitting results are statistically significant ($p \geq 0.1$) as defined by Clauset et al. (2009).

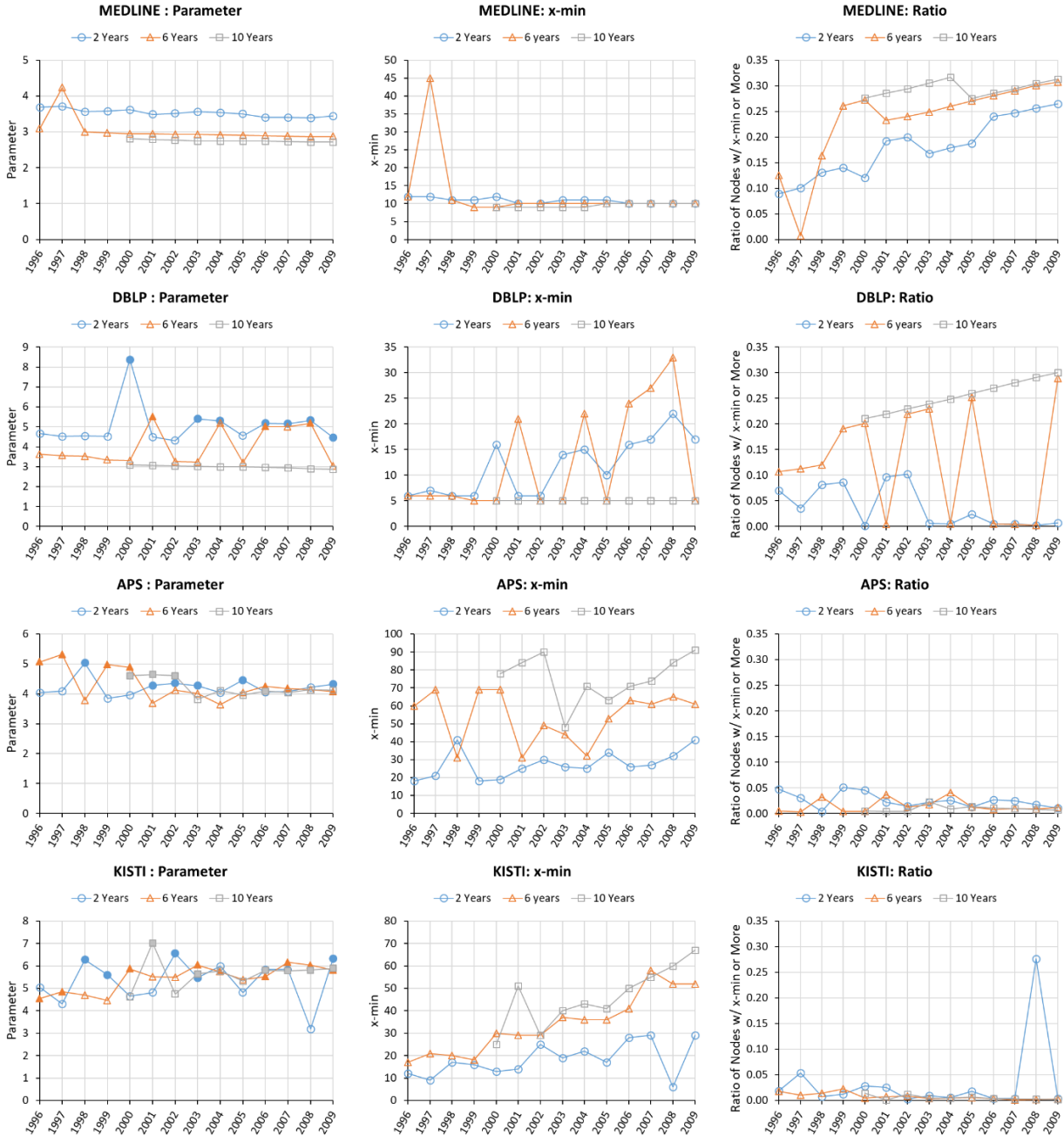

*Figure 10: Changes of Scaling Parameter, Minimum X (x-min), and Ratio of Nodes with X ≥ x-min (Filled shapes in left subfigures indicate statistically significant fitting results)*

According to Figure 10, degree distributions in DBLP, APS, and KISTI were fitted to different power-law slopes on different ranges of *x* values: data points in left (parameter) and middle (*x*-min) subfigures fluctuate with no distinct patterns, and many fitting results are not statistically significant (empty shapes). In contrast to that, parameter and x-min trends were quite stable in MEDLINE, but, as indicated by all empty shapes, all power-law fitting results in MEDLINE are not statistically significant. Even if we assume that all the fitting results were statistically significant (i.e., ignore filled or unfilled shapes), nodes with degrees *x-min* or more constitute at best 30% of all nodes in several networks spanning 6 and 10 years in MEDLINE and DBLP. In other networks, the ratios of nodes with *x*-min or more are very low:

mostly less than 5% in APS and KISTI, and less than 10% in DBLP. This implies that link formation involving small portions of nodes in our collaboration networks can be explained by preferential attachment models customized to power-law obeying degree distributions in empirical networks[12].

Conclusion and Discussion

Predicting links in collaboration networks is typically based on a process of identifying structural patterns of existing nodes that are present in a network but disconnected, and that form a link later on. Decent to high accuracy rates of link prediction based on this process have been achieved in prior studies (e.g., Guns & Rousseau, 2014; Liben-Nowell & Kleinberg, 2007; Yan & Guns, 2014) and a variety of computational solutions to this problem are available (e.g., Choudhury & Uddin, 2017, 2018; Guns, 2014). In this paper, we empirically identified methodological constraints to the eligibility of this process: considering links that involve at least one existing node (Type A, B, and C), only about 25% of all links in coauthorship networks adhere to the evolutionary process of disjoint nodes being present first and forming a tie later (Type B). Based on averaged empirical results from four datasets that represent different disciplines and a region, our findings show that roughly speaking, 25% of links represent continued collaborations (Type A), 25% of links are new collaboration between existing authors (Type B), and 50% are collaborations between a former and a new network member (Type C).

These observations do not imply that prior link prediction studies for collaboration networks are ineffective; rather, they show formational constraints that need to be overcome to increase the practical value of link prediction methods for collaboration networks. Specifically, the results of this study show that increasing the accuracy of computational solutions for link prediction may not be a reasonable goal when the underlying reality of link type distribution, i.e., the ratio of links that are eligible to the classic link prediction process, is the biggest roadblock to successful prediction. Even if a link predictor was 100% accurate, only half of the future links between existing nodes can be predicted (Type B out of Type A + Type B).

This roadblock cannot be easily removed with improved features, algorithms, or technology, but reflects the reality of link formation in scientific research collaboration. This indicates that link prediction solutions aiming at a high precision-recall rate for all links in networks would also need to account for a) sustained links (Type A), which is a computationally tractable task that can be solved with supervised machine learning techniques (Lü & Zhou, 2011; Mohdeb et al., 2016; Taskar et al., 2003), as well as links between (b) old and new network members (Type C), and (c) among new network members only (Type D). The latter two tasks (Type C and D) are difficult since the emergence of new members needs to be predicted first, which has insufficiently been addressed in link prediction research.

In addition, about half of all the links that involve an existing node contain one new node (Type C), which suggests that we should investigate this link formation process in more detail to be able to model it. Solving the problem might require us to reconsider several basic principles of how link prediction is done. Rather than attempting to find high performing predictors that are based on inherent network properties, we might need to incorporate external (outside of the network) factors such as node affiliation or homophily (Cabanac et al., 2015). This approach combined with network-structure-based models may lead to formulating high-performing models, which account for link formation of Type A and C as well as

---

[12] Many studies on power-law distribution in collaboration networks have fitted distribution tails (i.e., distribution of certain $x$ values and above) to power-law slopes to assess the performance of proposed network generation models. Several studies have divided a degree distribution into two parts (below and above a certain $x$ value) and fitted them separately to different power-law slopes (e.g., Wagner & Leydesdorff, 2005). A few others have tested power-law distributions with cut-offs (below certain $x$ value) (e.g., Newman, 2001b).

Type B. Our observations also provide a tangential insight into the social transaction costs of scientific co-authorship environments: many collaboration links involve two people who have not published together before (i.e., Type B and C). This means that for a large portion of collaboration relationships, coordination and trust need to be developed from scratch.

This study has several limitations. First, we analyzed co-authorship networks by following the research design used in several prior link prediction studies. Other social network datasets and analysis techniques may lead to different results about the ratio of sustained, partially new, and totally new links (e.g., Choudhury & Uddin, 2017, 2018). Second, the time slicing in our study followed common procedures, but was limited in the frame size (extended up to 10 years = past 5 years + present 5 years). Extending the window size of past versus present networks beyond this limit might provide different findings. Finally, relying on aggregated snapshots of past and present networks might not be a proper technique to trace link formation processes in networks. We would expect more nuanced insights if link formation would be traced across a whole dataset, not based on structural patterns, but on node identity. These are tasks for future research. Overall, link prediction for collaboration networks is in its early stages. We do not attempt to refute or negate prior studies. Instead, we hope our study provides new ideas for improving link prediction models for collaboration networks, and thereby contributes to this evolving research area.

## Acknowledgments

This work is supported, in part, by Korea Institute of Science and Technology Information (KISTI). We would like to thank Vetle Torvik (University of Illinois at Urbana-Champaign), the American Physical Society, DBLP, and KISTI for providing datasets. We are also grateful to Mark E. J. Newman (University of Michigan) for providing codes for disambiguating author names in APS data and Raf Guns (University of Antwerp) for comments on link prediction processes in *LinkPred*.

## Appendix

Degree Product: (Barabási et al., 2002) showed that if links in a network are formed based on preferential attachment, the probability of two nodes to form a link is proportional to the product of the degrees of those two nodes. This is frequently used to predict link formation among nodes present in both past and present networks. In the following equation, $S(x, y)$ is the prediction score for a pair of node x and y, and $\Gamma(x)$ is the set of nodes connected to x.

$$S(x, y) = |\Gamma(x)| \times |\Gamma(y)| \quad (2)$$

References

Adamic, L. A., & Adar, E. (2003). Friends and Neighbors on the Web. Social Networks, 25(3), 211-230. doi:10.1016/So378-8733(03)00009-1

Barabási, A. L., Jeong, H., Neda, Z., Ravasz, E., Schubert, A., & Vicsek, T. (2002). Evolution of the social network of scientific collaborations. Physica a-Statistical Mechanics and Its Applications, 311(3-4), 590-614. doi:10.1016/s0378-4371(02)00736-7

Braun, T., Glanzel, W., & Schubert, A. (2001). Publication and cooperation patterns of the authors of neuroscience journals. Scientometrics, 51(3), 499-510. doi:10.1023/A:1019643002560

Cabanac, G., Hubert, G., & Milard, B. (2015). Academic careers in Computer Science: continuance and transience of lifetime co-authorships. Scientometrics, 102(1), 135-150. doi:10.1007/s11192-014-1426-0

Chen, D.-B., Xiao, R., & Zeng, A. (2014). Predicting the evolution of spreading on complex networks. Scientific reports, 4.

Chen, H., Li, X., & Huang, Z. (2005, 7-11 June 2005). Link prediction approach to collaborative filtering. Paper presented at the Proceedings of the 5th ACM/IEEE-CS Joint Conference on Digital Libraries (JCDL '05).

Choudhury, N., & Uddin, S. (2017). *Mining Actor-level Structural and Neighborhood Evolution for Link Prediction in Dynamic Networks*. Paper presented at the Proceedings of the 2017 IEEE/ACM International Conference on Advances in Social Networks Analysis and Mining 2017, Sydney, Australia.

Choudhury, N., & Uddin, S. (2018). *Evolutionary Community Mining for Link Prediction in Dynamic Networks*. Paper presented at the Complex Networks & Their Applications VI, Lyon, France.

Clauset, A., Shalizi, C. R., & Newman, M. E. J. (2009). Power-law distributions in empirical data. Siam Review, 51(4), 661-703. doi:10.1137/070710111

De Nooy, W., Mrvar, A., & Batagelj, V. (2011). Exploratory social network analysis with Pajek (Vol. 27). New York: NY: Cambridge University Press.

Fegley, B. D., & Torvik, V. I. (2013). Has Large-Scale Named-Entity Network Analysis Been Resting on a Flawed Assumption? Plos One, 8(7), 1-16. doi:10.1371/journal.pone.0070299

Guns, R. (2014). Link Prediction. In Measuring Scholarly Impact (pp. 35-55): Springer.

Guns, R., & Rousseau, R. (2014). Recommending research collaborations using link prediction and random forest classifiers. Scientometrics, 101(2), 1461-1473. doi:10.1007/s11192-013-1228-9

Kim, J. (2018). Evaluating author name disambiguation for digital libraries: A case of DBLP. *Scientometrics, 116*(3), 1867-1886. doi:10.1007/s11192-018-2824-5

Kim, J., & Diesner, J. (2015). The effect of data pre-processing on understanding the evolution of collaboration networks. Journal of Informetrics, 9(1), 226-236. doi:10.1016/j.joi.2015.01.002

Kim, J., & Diesner, J. (2016). Distortive effects of initial-based name disambiguation on measurements of large-scale coauthorship networks. Journal of the Association for Information Science and Technology, 67(6), 1446-1461.

Kim, J., & Diesner, J. (2017). Over-time measurement of triadic closure in coauthorship networks. *Social Network Analysis and Mining, 7*(1), 1-12. doi:10.1007/s13278-017-0428-3

Kim, J., Tao, L., Lee, S.-H., & Diesner, J. (2016). Evolution and structure of scientific co-publishing network in Korea between 1948–2011. Scientometrics, 107(1), 27-41. doi:10.1007/s11192-016-1878-5

Lerchenmueller, M. J., & Sorenson, O. (2016). Author Disambiguation in PubMed: Evidence on the Precision and Recall of Author-ity among NIH-Funded Scientists. *PLOS ONE, 11*(7), e0158731.

Liben-Nowell, D., & Kleinberg, J. (2007). The link-prediction problem for social networks. Journal of the American Society for Information Science and Technology, 58(7), 1019-1031. doi:10.1002/asi.20591

Lü, L., & Zhou, T. (2011). Link prediction in complex networks: A survey. Physica A: Statistical Mechanics and its Applications, 390(6), 1150-1170.

Martin, T., Ball, B., Karrer, B., & Newman, M. E. J. (2013). Coauthorship and citation patterns in the Physical Review. Physical Review E, 88(1), 012814-012811~012819. doi:10.1103/PhysRevE.88.012814

Milojević, S. (2010). Modes of collaboration in modern science: Beyond power laws and preferential attachment. Journal of the American Society for Information Science and Technology, 61(7), 1410-1423. doi:10.1002/asi.21331

Mohdeb, D., Boubetra, A., & Charikhi, M. (2016). Tie persistence in academic social networks. Informatica, 40(3), 353.

Mollenhorst, G., Volker, B., & Flap, H. (2011). Shared contexts and triadic closure in core discussion networks. Social Networks, 33(4), 292-302. doi:10.1016/j.socnet.2011.09.001

Newman, D., Karimi, S., & Cavedon, L. (2009). Using Topic Models to Interpret MEDLINE's Medical Subject Headings. In A. Nicholson & X. Li (Eds.), AI 2009: Advances in Artificial Intelligence (Vol. 5866, pp. 270-279): Springer Berlin Heidelberg.

Newman, M. E. J. (2001a). Clustering and preferential attachment in growing networks. Physical Review E, 64(2). doi:10.1103/PhysRevE.64.025102

Newman, M. E. J. (2001b). The structure of scientific collaboration networks. Proceedings of the National Academy of Sciences of the United States of America, 98(2), 404-409. doi:10.1073/pnas.021544898

Pennock, D. M., Flake, G. W., Lawrence, S., Glover, E. J., & Giles, C. L. (2002). Winners don't take all: Characterizing the competition for links on the web. Proceedings of the National Academy of Sciences of the United States of America, 99(8), 5207-5211. doi:10.1073/pnas.032085699

Perc, M. (2014). The Matthew effect in empirical data. *Journal of The Royal Society Interface, 11*(98). doi:10.1098/rsif.2014.0378

Price, D., & Gürsey, S. (1976). Studies in scientometrics. 1. Transience and continuance in scientific authorship. Paper presented at the International Forum on Information and Documentation.

Reitz, F., & Hoffmann, O. (2011). Did They Notice? A Case-Study on the Community Contribution to Data Quality in DBLP. In S. Gradmann, F. Borri, C. Meghini, & H. Schuldt (Eds.), Research and Advanced Technology for Digital Libraries, Tpdl 2011 (Vol. 6966, pp. 204-215). Berlin: Springer-Verlag Berlin.

Resnick, P., & Varian, H. R. (1997). Recommender systems. Communications of the ACM, 40(3), 56-58.

Schubert, A., & Glanzel, W. (1991). Publication Dynamics - Models and Indicators. Scientometrics, 20(1), 317-331. doi:10.1007/Bf02018161

Taskar, B., Wong, M.-F., Abbeel, P., & Koller, D. (2003). Link prediction in relational data. Paper presented at the Advances in neural information processing systems.

Torvik, V. I., & Smalheiser, N. R. (2009). Author Name Disambiguation in MEDLINE. Acm Transactions on Knowledge Discovery from Data, 3(3), 1-29. doi:10.1145/1552303.1552304

Wagner, C. S., & Leydesdorff, L. (2005). Network structure, self-organization, and the growth of international collaboration in science. Research Policy, 34(10), 1608-1618. doi:10.1016/j.respol.2005.08.002

Yan, E., & Guns, R. (2014). Predicting and recommending collaborations: An author-, institution-, and country-level analysis. Journal of Informetrics, 8(2), 295-309. doi:10.1016/j.joi.2014.01.008